\documentclass[pdflatex,sn-mathphys]{sn-jnl}

\jyear{2022}%

\theoremstyle{thmstyleone}%
\theoremstyle{thmstyletwo}%

\theoremstyle{thmstylethree}%

\begin{document}

\title[Between News and History]{Between News and History: Identifying Networked Topics of Collective Attention on Wikipedia}


\author*[1]{\fnm{Patrick} \sur{Gildersleve}
\href{https://orcid.org/0000-0003-3714-6823}{\includegraphics[height=\fontcharht\font`\B]{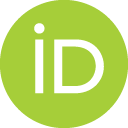}}
}\email{p.gildersleve@lse.ac.uk}

\author[2]{\fnm{Renaud} \sur{Lambiotte}
\href{https://orcid.org/0000-0002-0583-4595}{\includegraphics[height=\fontcharht\font`\B]{ORCID.png}}}\equalcont{These authors contributed equally to this work.}

\author[3]{\fnm{Taha} \sur{Yasseri} \href{https://orcid.org/0000-0002-1800-6094}{\includegraphics[height=\fontcharht\font`\B]{ORCID.png}}
}
\equalcont{These authors contributed equally to this work.}

\affil*[1]{\orgdiv{Department of Methodology}, \orgname{London School of Economics and Political Science}, \orgaddress{\street{Houghton Street}, \city{London}, \postcode{WC2A 2AE}, \country{UK}} \href{https://scholar.google.co.uk/citations?user=3CTTLDgAAAAJ}{https://scholar.google.co.uk/citations?user=3CTTLDgAAAAJ}}
\affil[2]{\orgdiv{Mathematical Institute}, \orgname{University of Oxford}, \orgaddress{\postcode{OX2 6GG}, \country{UK}} \href{https://scholar.google.co.uk/citations?user=lftXi28AAAAJ}{https://scholar.google.co.uk/citations?user=lftXi28AAAAJ}}
\affil[3]{\orgdiv{School of Sociology}, \orgname{University College Dublin}, \orgaddress{\postcode{Dublin 4}, \country{Ireland}} \href{https://scholar.google.co.uk/citations?user=AiLoMKwAAAAJ}{https://scholar.google.co.uk/citations?user=AiLoMKwAAAAJ}}

\abstract{The digital information landscape has introduced a new dimension to understanding how we collectively react to new information and preserve it at the societal level. This, together with the emergence of platforms such as Wikipedia, has challenged traditional views on the relationship between current events and historical accounts of events, with an ever-shrinking divide between ``news'' and ``history''. Wikipedia's place as the Internet's primary reference work thus poses the question of how it represents both traditional encyclopaedic knowledge and evolving important news stories. In other words, how is information on and attention towards current events integrated into the existing topical structures of Wikipedia? To address this we develop a temporal community detection approach towards topic detection that takes into account both short term dynamics of attention as well as long term article network structures. We apply this method to a dataset of one year of current events on Wikipedia to identify clusters distinct from those that would be found solely from page view time series correlations or static network structure. We are able to resolve the topics that more strongly reflect unfolding current events vs more established knowledge by the relative importance of collective attention dynamics vs link structures. We also offer important developments by identifying and describing the emergent topics on Wikipedia. This work provides a means of distinguishing how these information and attention clusters are related to Wikipedia’s twin faces of encyclopaedic knowledge and current events---crucial to understanding the production and consumption of knowledge in the digital age.}




\maketitle

\section{Introduction}
Since its founding in 2001, Wikipedia has grown from a simple, often dismissed, ‘Web 2.0’ based dream of shared knowledge to the Web’s primary authoritative information resource for billions of people. Its emergence has overhauled traditional conceptions of the encyclopaedia to become a rapidly updatable real-time record. In addition, it may act as an audience barometer for both historical knowledge and, crucially, current events. Wikipedia is thus an intriguing site of study for audience reception of news events---reflective of collective audience trends, yet somewhat divorced from the influence of journalistic and content delivery processes.

Wikipedia is a particularly attractive platform to analyse audience reception of current events. Beyond its position as one of the most popular sites on the World Wide Web, the information on Wikipedia is stored in a hyperlink network of articles. Information representation is then a result of both the knowledge structure on Wikipedia and the way its readers navigate that structure. This is unlike a news website, where there may be a single news article, or occasionally a small series of articles towards each event, often disconnected from each other, despite sharing similar subjects. Wikipedia data then facilitates connections and comparisons between different events, as well as to longer term historical knowledge structures. Moreover, using Wikipedia as a data source to study current events fulfils calls for the use of ``extra-media data''---information about current events not taken directly from news media \cite{rosengren1970international}. Various works have taken advantage of Wikipedia's online importance, relevance towards current events, and data availability, using different aspects of the real-time digital encyclopaedia in studying areas such as stock prices \cite{moat2013quantifying}, film box office performance \cite{mestyan2013early}, and disease spread \cite{mciver2014wikipedia}.

In order to holistically study the interaction of current events with Wikipedia, we must first identify what events are represented and how they are manifested. Individuals browse groups of articles relating to events in the news, but what are the various groups that are browsed? Are they consistent between events in forming news topics? And how well do access patterns align with the structure of the information in the encyclopaedia? One must answer these questions in order to study how different kinds of events are integrated into collective memory, encyclopaedic knowledge, and the historical record.

Previous research has made progress in exploring different modes of data, such as page views, hyperlinks, and edits \cite{Georgescu2013, ahn2011wikitopics}, as well as focusing on specific events or event categories, such as natural disasters or sporting events \cite{Keegan2011, kobayashi2021modeling}. However, these efforts have often been limited in scope and/or have not been able to connect directly with news records \cite{miz2019anomaly, miz2020trending}. Therefore, there is a need for comprehensive techniques that can link specific events, integrate page view attention patterns with hyperlink network structure, and extrapolate individual events to broader topics, all simultaneously.

To address this gap, in this work we present a topic detection model for Wikipedia. Our method leverages both Wikipedia hyperlink network structure and correlations between page view time series, combined with a database of events from Wikipedia's current events portal, to identify groups of articles that are both well connected \emph{and} exhibit similar patterns of page views around individual events. These groups of articles are then connected through time to identify the recurrent topics attracting attention on Wikipedia. The resulting `Event Reactions' and `Topics of Attention' encompass news topics as well as attention towards background topics on Wikipedia (and can be resolved as such). These objects are revealing of various interesting features on the nature of knowledge and news recorded on Wikipedia and can act as general purpose event and topic summaries for study in future work. Our work is built around the following research questions.

\begin{itemize}
    \item \textbf{RQ:} How are current events represented in the knowledge structures and access patterns of Wikipedia and its users?
    \begin{itemize}
        \item RQa: Can we meaningfully and robustly sample the related groups of Wikipedia articles associated with a given news event?
        \item RQb: Are these groups of articles from different events related and do they form coherent topics?
    \end{itemize}
\end{itemize}

We find that our approach yields communities of articles (`Event Reactions') that are strongly related according to both hyperlink network structure and correlated page views. When considering Event Reactions' relationships to each other we find that they form coherent, human-validated, higher-level `Topics of Attention'. These topics may be resolved to volatile news topics and stable background topics, representing both facets of Wikipedia as a stable knowledge base and rapidly updating current events record. The topics themselves also qualitatively exhibit a background historical concept space, strong geographical effects, a focus on individuals, and breakout subtopics.

\section{Related Work}

Historic variation in news and encylopaedism is based on the discrepancy in information distribution methods. The Internet has narrowed, even brought about a convergence in the time, format, and audience for these two initially very different knowledge exchange media. Through Wikipedia the encyclopaedia as a format has come to be a highly responsive trove of information on current events, plus readily accessible online news archives (occasionally from citizens' records) comprehensively chart the events of the 21st century. News and encyclopaedic recording, together with the public's experience of them, have never been more similar. This convergence has been thoroughly explored \cite{keegan2012high, gildersleve2021wikipedia}, and is also reflected in reckonings on its significance to ``open-source history'' \cite{rosenzweig2006can}, and collective memory \cite{pentzold2009, Ferron2011, kanhabua2014, luyt2015, garcia2017memory, twyman2017black, candia2019universal, yasseri2022collective}. Issues of contestation and distortion of this prominent digital record of collective memory are also attracting increasing attention, for example, the cases of widespread administrator-driven far-right historical revisionism on Croatian Wikipedia \cite{wikipedia_2021Cro} and distortion of Holocaust history on the English edition \cite{grabowski2023wikipedia}.

Theoretical work on news media has long had to wrestle with the fact that news outlets themselves, in (justifiably) selectively covering events, are not necessarily representative of current events more widely or the audiences' views on them. Karl Erik Rosengren calls for studies that incorporate ``extra-media data''---information about current events not taken directly from news media---to help address this issue \cite{rosengren1970international}. In the Internet age, we may now turn to the large scale tracking of user access patterns and active constructions of repositories of knowledge for a fresh perspective on this extra-media data.

Wikipedia is of course not a news website. It is, however, used as a secondary information resource, which individuals use to further research and contribute towards topics they have encountered through other news media. According to respondents to Wikimedia surveys, 13\% of readers visit the site directly because of current events, and a further 30\% visit due to wider media coverage\footnote{No noted percentage for usage of Wikipedia as a news source.}\cite{Singer2017}. Perceptions around the reliability of Wikipedia have only improved since its early days, teachers' and lecturers' warnings have grown increasingly futile, with various studies confirming its overall reliability on a range of subjects \cite{giles2005internet, devgan2007wiki, fallis2008toward, messner2011legitimizing, messner2013wikipedia}. From bar wager to academic papers \cite{thompson2018science}, its authority as the unofficial arbiter of social facts is undeniable. As such, Wikipedia is an appealing representative site for predicting external collective behaviour, both on and offline, from search trends \cite{yoshida2015wikipedia}, to stock and cryptocurrency prices \cite{moat2013quantifying, elbahrawy2019wikipedia}, film box office performance \cite{mestyan2013early}, city tourism numbers \cite{hinnosaar2021wikipedia}, disease spread \cite{mciver2014wikipedia}, and election results \cite{yasseri2014, yasseri2016}.

Moreover, plenty of web companies rely on Wikipedia's content in powering their own services. For example, platforms such as Facebook, Google, YouTube, Twitter, Amazon Alexa, and Apple Siri use Wikipedia in producing their own knowledge graphs, informing automated search results and infoboxes, verifying of notable persons, directing their users to authoritative sources on issues of conspiracies and misinformation, and in building powerful large language models \cite{matsakis_2018, withers_2018, perez_2020, twitter_2020, vincent2021deeper, thoppilan2022lamda}. The collective telling of events from Wikipedia, and the aggregate user behaviour in browsing it, is thus emblematic of the kind of audience-centric extra-media data required for studying news media.

Studying the news through Wikipedia article data can be considered a form of content analysis, a common technique in communication and journalism studies. Of particular interest is the task of automatic content analysis, whereby topics, trends, agenda, etc. are analysed across large corpora of news stories where manual coding is not feasible. Three main forms of this are identified in \cite{grimmer2013text} and \cite{boumans2016taking}; rudimentary dictionary based methods (e.g. \cite{van2017economic}), supervised machine learning approaches (e.g. \cite{scharkow2013thematic}), and unsupervised machine learning approaches (e.g. \cite{guo2016big}). A novel unsupervised approach in which individual news stories are clustered into `news story chains' according to textual similarity (particularly with novel words), grouping individual articles and their follow-ups into single entities is offered by \cite{nicholls2019understanding}.

A key feature of Wikipedia that separates it from traditional encyclopaedias and news media (as well as in truth much of the modern social web) is the fact that content on Wikipedia exists in a single hyperlinked network of articles. What is constructed by different groups of editors, as well as how it is built, better informs us of the information itself. Prior literature has explored how these cultural differences manifest on the site \cite{Aragon2011}, how the network itself drives rich collaborative environments \cite{kane2009s}, as well as how related knowledge graphs can be used for automated fact checking \cite{Ciampaglia2015}. Previous literature typically concerns itself with events that have dedicated articles \cite{Osborne2012, Garcia-Gavilanes2016}. However, news events can also be documented within one already existing article or across several different articles. For example, stories about one public figure making newsworthy comments about another may be separately recorded on their respective articles.

Work on the concept of spillovers is also relevant here. Even in cases where content in linked articles is not directly relevant to the focus of some exogenous shock, attention in page views and edits may still `spillover' to a selection of neighbours. This has been studied in the context of current events \cite{Kummer2014, garcia2017memory}, but also the effects of articles being featured on the Wikipedia main page \cite{kummer2013spillovers}, or editing campaigns \cite{zhu2020content}. Analysing the dynamics of news events in the context of their links to related topics thus clearly necessitates a network based approach.

Several papers cover methods for the detection and summarisation of news events using activity on the platform, frequently relying on bursts in edits \cite{Georgescu2013} or page views \cite{ahn2011wikitopics, Miz2017}, or alternatively dedicated databases for particular event categories \cite{kobayashi2021modeling}. Collecting and comparing events from a common category can be informative in answering how event dynamics vary according to category specific parameters. However, further work is needed in comparing across categories. Efforts have also been made to enhance news event detection by combining Wikipedia with other forms of social media acting as a source \cite{Osborne2012} or filter \cite{Steiner2013} for events. Bursts of attention towards a topic seem like a natural way of selecting news events but clearly do not cover the full expressible range of event dynamics. Case studies considering edit and page view dynamics \cite{Keegan2011} or work studying breaking news events more generally \cite{Keegan2013} look to address how users' collective attention towards articles situated in a wider topic network evolves, and how it can drive the collaborative editing activities that shape the content and structure of knowledge on the website. News events are rapidly covered on Wikipedia, yet they can have a lasting impact on article content and network structure, incrementally contributing to the collective knowledge base of Wikipedia.

In doing so, individual events shape and integrate into the wider topics represented on Wikipedia. Work that approaches the task of identifying and summarising these topics can use semantic information \cite{ni2009mining, ni2011cross}, the article network structure \cite{syed2008wikipedia, Miz2017}, category tags \cite{kittur2009s, boldi2016cleansing}, and page view patterns \cite{Miz2017, miz2019anomaly, miz2020trending}. Several of these approaches are also language agnostic \cite{Miz2017, miz2019anomaly, miz2020trending, johnson2021language}, or even multi-lingual \cite{ni2009mining, ni2011cross}. Most notable of these approaches is that of \cite{Miz2017, miz2019anomaly, miz2020trending} whose language agnostic community detection model incorporates correlated page views together with article network structure. However, in cases where larger datasets are used (such as the topic-level analyses) it is frequently the case that properties such as page views are studied independent of any explanatory description, with any detected interesting features such as peaks later being ascribed meaning by the researcher(s) (likely to some external event) \cite{lehmann2012dynamical}. There is an important distinction between starting from the point of current events and understanding their dynamics, rather than observing particular dynamics (such as bursts and anomalies) and later attempting to relate them to news events. Firstly, sampling-wise we may only select for particular dynamics when adopting the latter approach (as already touched upon). Secondly, immediately linking to news events assists in later stages in the interpretation of results.

\section{Wikipedia Data}
\label{sec:data}

The three primary classes of Wikipedia data used are information on events that occur from the Wikipedia Current Events Portal \cite{current_events_2019}, data for the article network of Wikipedia (i.e., the article names and what hyperlinks exist between them), and time series data for the daily page views to each article. Supplementary data on Wikipedia redirects is also used. Further detail on data and how it is obtained may be found in Appendix \ref{appendix:data}. All code and data is available through the WikiNewsTopics GitHub repository\footnote{\url{https://github.com/pgilders/WikiNewsNetwork-01-WikiNewsTopics}}.

\subsection{Current Events Portal}

\begin{figure}[h]
\centering
\includegraphics[width=\textwidth]{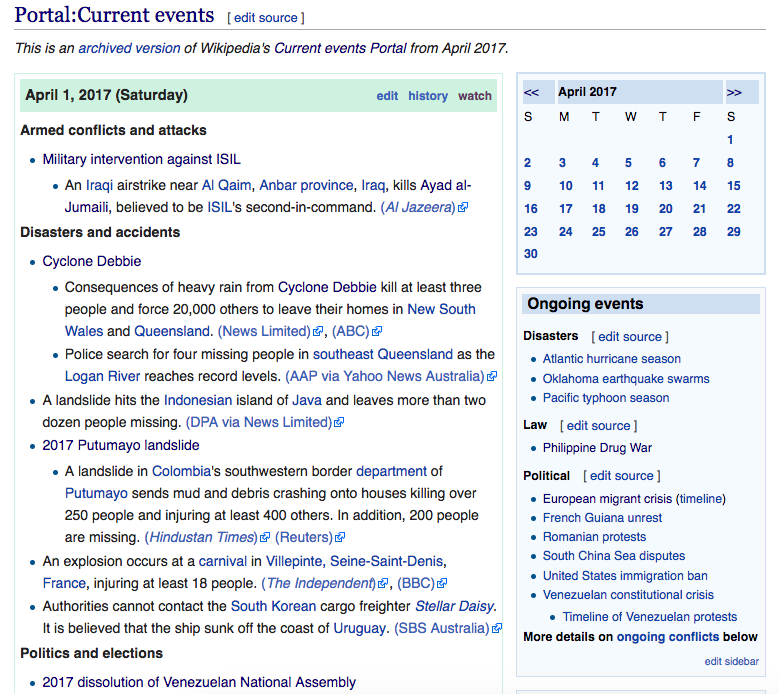}
\caption{A snapshot of the Wikipedia current events portal. Live version available at \url{en.wikipedia.org/wiki/Portal:Current_events}.}
\label{fig:portal}
\end{figure}

The Wikipedia Current Events Portal is a daily archive of events as recorded by Wikipedia editors. Whilst records are in English, coverage is (nominally) of global events of international interest. Events are sorted in 10 categories and are recorded with a summary sentence with links to relevant articles. A partial snapshot is displayed in Figure \ref{fig:portal}. We scrape the page to sample a full year of events from 1\textsuperscript{st} December 2017 to 30\textsuperscript{th} November 2018. Initial data gathered for each event includes the date, category, full text description, and the linked Wikipedia articles (henceforth referred to as ``core articles'') in each description. For example, in Figure \ref{fig:portal}, the final event on 01/04/2017 in the `Disasters and Accidents' category is described as ``Authorities cannot contact the \textit{South Korean} cargo freighter \textit{Stellar Daisy}. It is believed that the ship sunk off the coast of \textit{Uruguay}''. We extract the linked pages \textit{South Korea} (displayed text does not have to match article title), \textit{Stellar Daisy}, and \textit{Uruguay} as the core articles for this event. In total, 7919 events are gathered from this year-long period.

\subsection{Clickstream Networks}

For the Wikipedia network data, we use dumps from the Wikipedia Clickstream \cite{clickstream}. The Wikipedia Clickstream contains monthly aggregated counts for the number of times links are accessed on Wikipedia, and crucially where from, in (referrer, resource)---equivalently (source, target)---pairs. Only hyperlinks which are clicked more than 10 times in a month are included in the dataset. We only include hyperlinks between Wikipedia articles, excluding links from external to Wikipedia and from Wikipedia's Main Page. Essentially, this represents an edgelist that forms a directed, weighted network of monthly navigation between Wikipedia articles. Clickstream data for the English Wikipedia was downloaded for November 2017--December 2018, allowing a one month buffer for studying events at the start and end of the time period of study.

\begin{table*}[h]
\centering
\resizebox{\textwidth}{!}{
\begin{tabular}{l|lllll}
 & Timespan & Resolution & Source & Redirects (raw data) & Notes \\ \midrule
Current Events Portal & 12/17-11/18 & Daily   & Scraped    & Unresolved & Categorised by news category      \\
Clickstream Networks  & 11/17-12/18 & Monthly & Data Dumps & Resolved   & 10 click threshold for each link (edge)           \\
Page View Time Series & 11/17-12/18 & Hourly  & Data Dumps & Unresolved & Separate mobile \& desktop page views
\end{tabular}
}

\caption{A summary of the Wikipedia data used. Further detail available in Appendix \ref{appendix:data}.}
\label{tab:dataD}
\end{table*}

\subsection{Page View Time Series}

Page view data in hourly granularity for all articles is downloaded from the Wikimedia data dumps \cite{wikimediadumps}. We identify the networks of articles linked to the entries in the Current Events Portal for which page view time series are required (more details in Section \ref{sec:CEprocess}), and process the raw compressed time series data to more accessible HDF5 format. This data was downloaded for the period November 2017--December 2018. 

\subsection{Redirects}

Wikipedia article redirects are not resolved in all data sources. However, we must account for their important role in Wikipedia's structure and the shaping of traffic \cite{hill2014consider}. Wikimedia API calls for redirects \cite{wiki:api} were used to create a mapping using 1) what `correct' article title they redirect to (if necessary) 2) all other names that redirect to the article. Page views for individual articles were then calculated by summing those for the groups of their redirects. When mapping the page view data to the articles in the clickstream data this guarantees correct correspondence. The different forms of data are summarised in Table \ref{tab:dataD}.

\section{Methods}
\subsection{Building Event Networks}
\label{sec:sampnews}
\label{sec:CEprocess}

Here we detail the exact pipeline by which we generate a network of related articles and associated page view time series related to each event (the `Event Networks'), analyse these for communities of articles representing distinct content and dynamic based `Event Reactions', and finally cluster these communities (the `Event Reactions') based on overlapping constituent Wikipedia articles to identify `Topics of Attention'. These concepts are the key levels of analysis in this work, and are more clearly defined as follows:

\begin{itemize}
    \item \textbf{Event Network:} The hyperlink network of Wikipedia articles and associated page view time series related to a particular news event.
    \item \textbf{Event Reaction:} A community of articles within a single Event Network that are relatively strongly linked and receive correlated patterns of page views.
    \item \textbf{Topic of Attention:} A cluster of Event Reactions from different events, grouped according to common constituent Wikipedia articles.
\end{itemize}

Firstly, entries from the Current Events Portal must be related to networks of Wikipedia articles and page view data. The process to generate `Event Networks' runs as follows:

\begin{itemize}
    \item For each event:
    \begin{itemize}
    \item Scrape event data from Current Events Portal.
    \item Resolve redirects of `core articles' linked in event description.
    \item Use clickstream data to create network of all articles that link to, and are linked to by, as well as all links between these articles over a window of 61 days centred on the recorded event date. Edge weights are a weighted average of the monthly click totals (weights based on fraction of 61 day window in each month).
    \item Keep all edges with weight $>100$ (i.e. removing edges clicked infrequently), remove any isolates. This is done primarily due to computational speed and memory limitations regarding the sparsity of graphs during later community detection stages.
\end{itemize}
    \item Collect all article names in the networks, with all redirects, and the period of time they are in the news and require page view data for.
    \item Process the page view data, keeping data for all required article names, with redirects, over the required time periods.
    \item Assign 61-day time series (30 days before/after event date) to each event for each article in the respective networks.
\end{itemize}

The Event Networks encapsulate the network structure and page view dynamics in both anticipation and response to current events, however, they are not a wholly satisfactory description when attempting to generate summary statistics for each event. It is not the case, given the breadth of pages included in each network, that all their articles will exhibit the same signals for page views or edits, many simply being unrelated in the context of the news event. As such, simple averaging techniques for network level features will likely wash out any useful information.


\subsection{From Event Networks to Event Reactions}

One might think that the issues with the Event Networks are a result of the sampling strategy. In abstract terms, there may well be one true signal for each news event, yet it is obfuscated by the noise of less related pages picked up in the network, or concurrent news events involving the same pages, and that the solution is simply some filter or averaging process. We argue that on the contrary it is the very nature of current events that when studying the underlying constituent concepts there may be a variety of responses. These may be due to longer term effects from historical events, structural effects from related information, as well as associations with other current events. This could lead to a variety of different page view patterns. This may seem trivial, but it often does not explicitly emerge in research where the objects of study in focus are specific individual hashtags, news articles, YouTube videos, etc.

The constituent Wikipedia articles relating to individual events exhibit a variety of different dynamics tied to historical, structural, and concurrent news effects. We thus propose a method to separate responses across both content (structure) and attention (dynamics), to identify which groups of articles are both well connected \textit{and} exhibit similar page view time series. Simply taking clusters according to network structure ignores short term associations. On the other hand, simply taking pages with correlated responses ignores context of the related content, and could also introduce spurious associations. This approach takes both factors into account.

\begin{figure}[h]
\centering
\includegraphics[width=\textwidth]{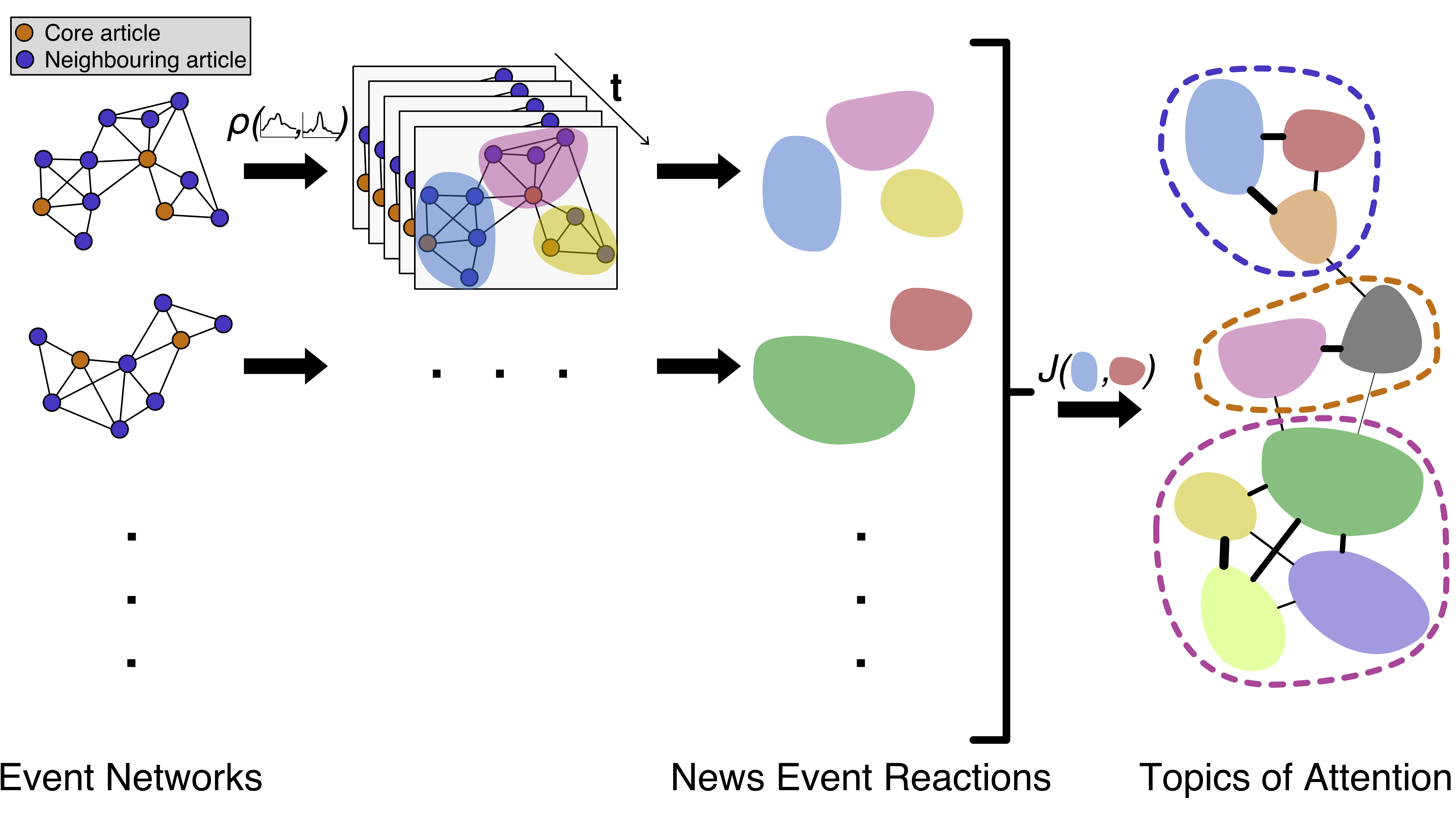}
\caption{A schematic of the processing of Event Networks to Event Reactions to Topics of Attention. For each Event Network, edge weights from rolling page view Pearson correlations ($\rho$) are calculated between article time series. Temporal community detection extracts the Event Reactions which (through Jaccard similarity $J$) form the higher level-network. We then identify the Topics of Attention with another stage of community detection.}
\label{fig:schematic}
\end{figure}

The chosen two-stage temporal community detection approach disentangles the different response signals across each Event Network into communities of articles termed `Event Reactions'. Each news event is partitioned into a handful of `Event Reactions' across the different subjects represented. These `Event Reactions' are then clustered with those from other news events according to common constituent articles (via a weighted Jaccard index), to detect broader topics, termed `Topics of Attention'. A schematic of the full process is shown in Figure \ref{fig:schematic}\footnote{The reader familiar with topic modelling may find the following analogy useful. Individual Event Networks represent `documents', that are made up of Wikipedia articles, akin to `words', and each news event represents a sample of wider news `topics', that we term `Topics of Attention'.}. 

\subsection{Temporal Community Detection of Signals on Knowledge Networks}
\label{sec:Gi}

Together with Section \ref{sec:sampnews}, here we tackle \textbf{RQa:} \textit{Can we meaningfully and robustly sample the related groups of Wikipedia articles associated with a given news event?} For a given news event $i$, with associated network of articles $G_{i}(V, E)$ (nodes representing articles, edges representing hyperlinks between them), there is an associated set of time series for page views towards articles $p_n(t) \forall n \in V$ with $T$ timesteps. To measure similarity in patterns of collective attention towards articles, we calculate Pearson correlations of these time series for all linked nodes over a rolling 7 day window, yielding $W_{\text{edges}}$, which is a $\|V\|\times\|V\|\times(T-6)$ dimensional array. The full correlation matrix would be represented by $W$, so that $W_{\text{edges}} = W \circ (A + A^T)/2$, where $A$ is the (unweighted) adjacency matrix of $G_i$ and $\circ$ denotes the Hadamard product between the matrices. We exploit the information in the sparsity of the hyperlink network to only calculate correlations for a tiny fraction of node combinations, allowing the algorithm to scale well to large networks. The operation also enriches the static network structure with temporal information. Related approaches could find use in other domains, e.g., in neuroscience to combine structural network information (DTI) with time series (fMRI) \cite{hagmann2008mapping, michel2018eeg}.

$W_{\text{edges}}$ represents an undirected (note that Pearson correlation is a symmetric measure, which motivates the removal of directedness), weighted temporal network $G'_i(t)(V, E(t))$, on which we perform community detection to identify groups of articles that are both well connected by hyperlinks, and exhibit correlated patterns of page views.

The Leiden algorithm \cite{traag2019leiden} is selected for temporal community detection. This is an extension of the popular Louvain algorithm \cite{blondel2008fast}, but addresses an issue whereby communities may be arbitrarily badly connected, it also runs faster than the Louvain algorithm. Rather than standard modularity, the Constant Potts Model \cite{traag2011cpm} is used for the quality function, since it can handle both positive and negative edge weights (which can in principle be observed), the readily interpretable resolution parameter, and the independence of communities from the observed graph/subgraph (particularly important given the articles present are a sample of the much larger Wikipedia article network). To extract information from the temporal network, we further adapt the method proposed in \cite{mucha2010community}, by considering the $T-6$ layers of the temporal graph, and connecting the same node in successive layers by an interlayer edge with weight $\tau=1$. By doing so, the temporal network is represented as a static, weighted network where each node appears $T-6$ times. We optimise its Constant Potts Model with the Leiden algorithm, thereby uncovering communities that are made of nodes at multiple times. Note that this operation requires the determination of the resolution parameter. A search for this parameter with a robustness test on a 100 event sample is carried out in Appendix \ref{appendix:robustness1}, with the resolution being set to $r=0.25$.

For each Event Network, the obtained partition $\mathbf{P}_i$ is comprised of a handful of communities $C_{ij}$. Any detected communities which contain at least one of the `core articles' from the descriptive text of the event, and that overlap in time with the day of the event are kept as Event Reactions---$R_{ij}$. Each of these elements is in effect a building block of wider Topics of Attention. The discrepancy in timescales between fast-pace attention towards news events and the more slowly evolving structure of the Wikipedia article network means these topics are not necessarily reflected in solely the hyperlink structure, or solely through correlated short term page views. In addition, satisfactory temporal community detection on one network for one year over the $\approx6$ million English Wikipedia articles is not computationally feasible. From the 7,919 events, we obtain 7,823 Event Networks with more than 1 node and edge (since a small number of event records do not have a popular associated article), and generate 26,579 Event Reactions.

\subsection{Community Detection Comparison}
\label{sec:cdc}
\subsubsection*{Capturing excess page views}
Our objective is to collect as good a selection as possible of Wikipedia articles, representative of a particular event. We expect articles related to some current event will exhibit a heightened level of page views around the time of the event. To describe the dynamics of attention around the event, we should select the communities containing core articles that on average exhibit excess page views around the time of the event. Communities that do not contain core articles are deemed not relevant to the event---the constituent articles are not structurally connected well enough to the core articles and/or page views do not follow a similar enough pattern. In cases where a community contains a core article but does not exhibit an increase in page views, we can conclude the core article, and the rest of the community, are background articles not directly relevant to the event, and can put them aside when focusing on event page view dynamics. We can then measure how well we have captured the event with the total excess page views towards articles in the identified relevant communities. We can compare how different community detection approaches perform on this measure on each event $i$, defined as 
\begin{equation}
     \text{Excess}_i = \sum_{\substack{R_{ij} \in \mathbf{P_i},\\ \max_{-1 \leq t \leq 1}(\widetilde{q}_{ij}(t)) > 3}} \sum_{k \in R_{ij}} \sum_{t'=0}^{t'=6}{p_k(t') - \text{median}(p_k(t))}.
\end{equation}
Here $R_{ij}$ refers to an Event Reaction (community containing a core article) in partition $\mathbf{P_i}$, $\widetilde{q}_{ij}(t)$ is the total page views towards articles in Event Reaction $j$, centred on the median value and scaled to the interquartile range, and $p_k(t)$ is the page view time series of article $k$ in Event Reaction $j$. Effectively, by the condition on the first summation, we only consider communities that contain a core article, and that have increased overall views around the time of the event. For each of these selected communities, we sum the page views above the median values over the first 7 days for each of the constituent articles.

\begin{figure}[]
    \centering
    \includegraphics[width=\textwidth]{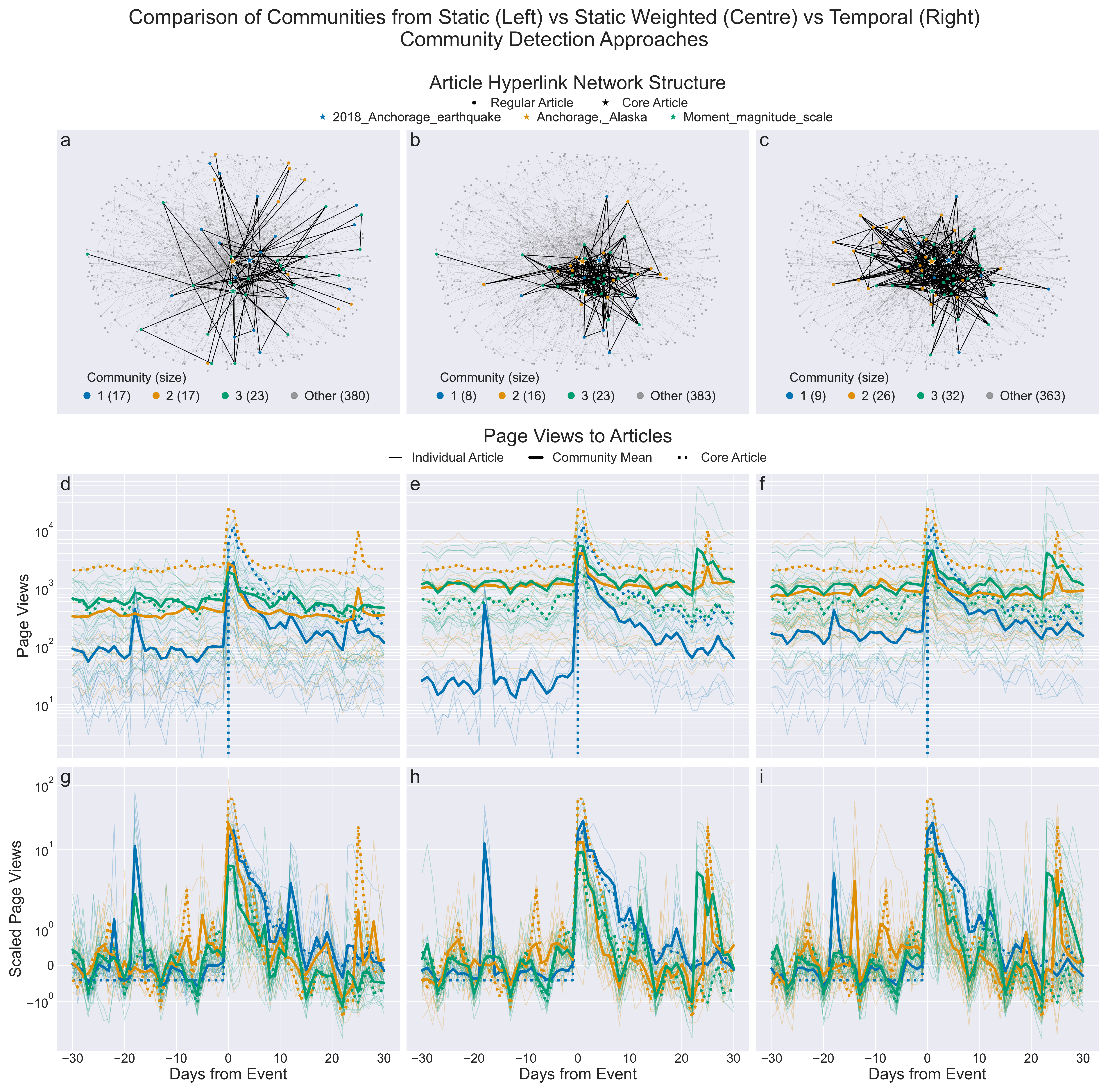}
    \caption{A comparison between the communities obtained through different methods. The event record in question is \textit{``2018/11/30. [[2018 Anchorage earthquake]]: A [[magnitude]] 7.0 earthquake hits Alaska, with the epicenter in [[Anchorage]]. Severe damage is reported.''} (core articles indicated by square brackets). In all three approaches there is a community centred around each of the core articles `2018 Anchorage earthquake' (a new article dedicated to the event), `Anchorage, Alaska', and `Moment magnitude scale'. The structural, navigational, and temporal approaches capture 263,585, 489,538, and 546,978 excess views, respectively. The absolute page views (in terms of total and mean) increase with the temporal approach (f vs d\&e), yet the scaled page view patterns remain similar (i vs g\&h). This indicates that a number of additional articles with similar spikes in attention relating to the earthquake have been identified. These articles were not captured in any of the previous static communities. With a static approach, in only taking the articles that are structurally/navigationally close to the “core” articles, we may both miss where attention is being directed by this event and markedly underestimate the amount of attention towards it.}
    \label{fig:exampleevent}
\end{figure}

To make an instructive comparison, we can consider how community detection approaches with just structural hyperlink information, aggregate navigational information, and our method perform on the excess views measure. For each event, we calculate the captured excess views in communities from our temporal approach against that from a simple implementation of the Leiden algorithm on both the static, unweighted, ``structural'' hyperlink network, as well as against against a static, weighted, ``navigational'' network, where edge weights are set to the weighted average link clicks from the clickstream data (as used in Section \ref{sec:CEprocess}). In both cases, resolutions were selected in a similar fashion to as in the temporal approach, detailed in Appendix \ref{appendix:robustness1}. Comparing the results on the captured excess page views, the temporal approach captures at least as many excess page views as the structural approach in 72.4\% of events and at least as many excess page views as the navigational approach in 72.8\% of events. Taking the ratios of captured excess views ($\text{Excess}_i^{\text{Temp}}/\text{Excess}_i^{\text{Struc}}$ and $\text{Excess}_i^{\text{Temp}}/\text{Excess}_i^{\text{Nav}}$) and considering the geometric mean across all events, the temporal approach captures 1.13 times the excess views of the structural approach and 1.23 times that of the navigational approach. Taking the median, the temporal approach captures 1.04 times the excess views of both the structural and navigational approaches. Our method better captures articles in communities relevant to current events. An example comparing the three approaches on a single event is provided in Figure \ref{fig:exampleevent}.

\subsubsection*{Structural similarity}

A follow-up task is to compare the makeup of the obtained communities from our combined network structure and page view approach correlations against those generated from a solely network structure approach. If the communities from the combined approach are no different from those for the network structure baseline, then this indicates attention dynamics in response to any news event have very little effect, and the community is well represented solely by the network. Any `disturbance' by the news event is either minimal, or closely aligns with how information is already represented on Wikipedia. On the other hand, if the communities obtained from each approach are quite different, then the variation in page view dynamics among the articles in the network is important in producing the Event Reactions. Any `disturbance' by the news event is of sufficient magnitude that the association between concepts related to the event is then not well represented by the relatively static network structure on Wikipedia.

The approach for each event is based on comparing the communities already obtained from the temporal network of correlated time series to those we obtain from community detection on a single-layer, unweighted graph representing solely the structure of the article network, without user attention and navigation patterns, (i.e. $G_i$ from Section \ref{sec:Gi}). For each event $i$ we run community detection on the graph $G_i$ over the same logarithmic range of resolutions $r \in [1.23\times10^{-4}, 1]$ from the robustness tests (Appendix \ref{appendix:robustness1}), yielding the partitions $\mathbf{P}'_{ir}$. Each Event Reaction from the temporal approach ($R_{ij}$) receives a `Structural Similarity' score $s_{ij}$. This score is defined as the maximum of the similarities between $R_{ij}$ and each community obtained from the non-temporal approach across all resolutions $C'_{ikr} \in \mathbf{P}'_{ir}\ : \  1.23\times10^{-4} \leq r \le 1$. Thus,
\begin{equation}
    s_{ij} = max(J_w(R_{ij}, C'_{ikr})\ \forall \  C'_{ikr} \in \mathbf{P}'_{ir}\  : \  1.23\times10^{-4} \leq r  \leq 1),
\end{equation}
where $J_w(x, y)$ is the Jaccard similarity between communities $x$ and $y$, weighted by the PageRank scores \cite{page1999pagerank} of nodes in the subgraphs $x$ and $y$ \cite{ioffe2010improved} (weighting more important articles to the community more highly). This takes into account both the content of each community in terms of Wikipedia articles, and the relative importance of said articles within their respective communities.

\begin{figure}[t]
    \centering
    \includegraphics[width=\textwidth]{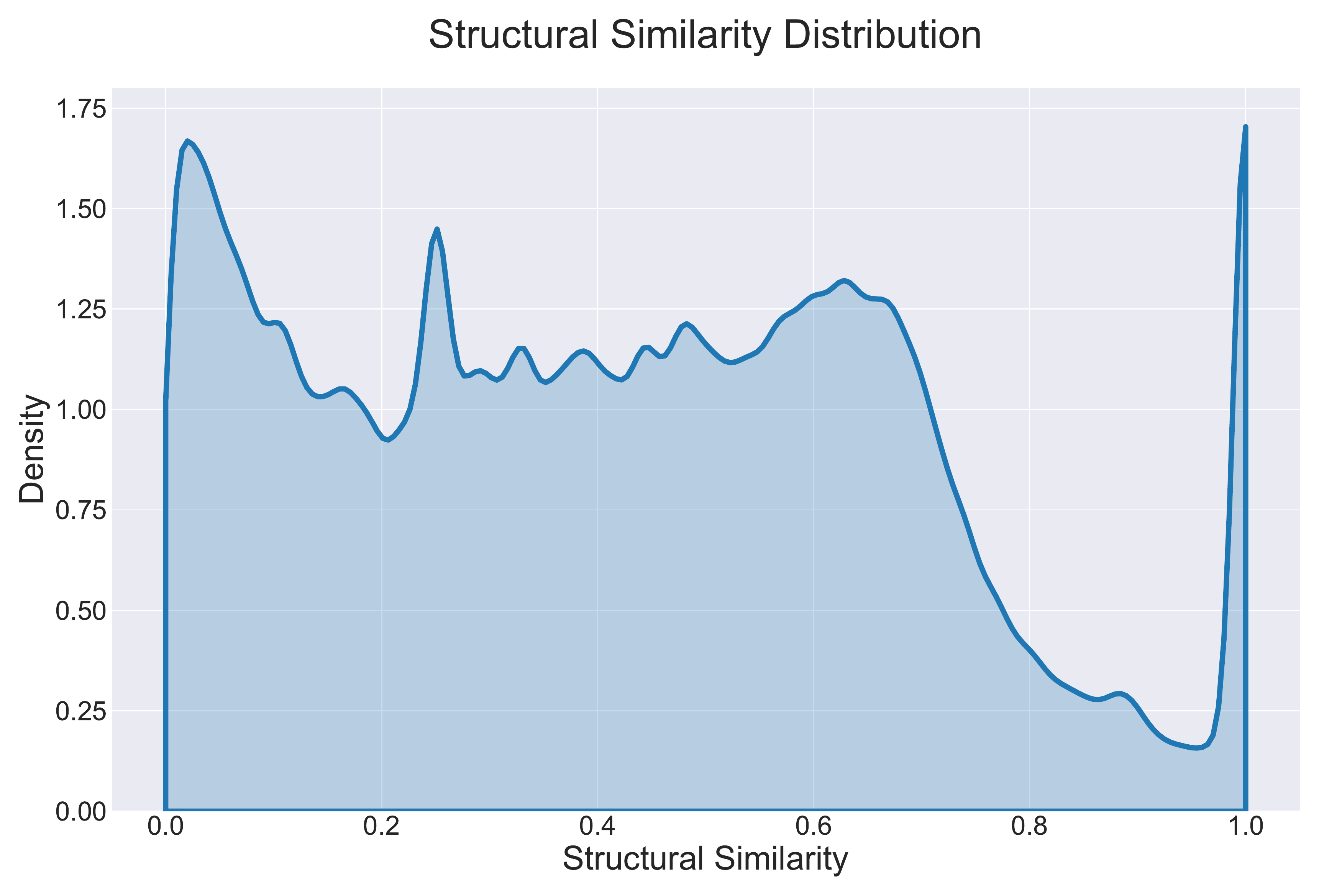}
    \caption{Distribution for the structural similarity scores of all Event Reactions.}
    \label{fig:SSD}
\end{figure}

The Structural Similarity score describes how dependent the observed community $R_{ij}$ is on variation in short term correlated attention dynamics in an Event Network, compared to the longer term network structure. If all page view time series were uniformly correlated, we would expect $s_{ij} \approx 1$, i.e., all edge weights would be approximately equal, and the community detection is more reliant on the presence/absence of edges. If on the other hand a subset of articles receive strongly correlated page views, uncorrelated with the page views to other articles, we would expect $s_{ij} \approx 0$, i.e., community detection is more dependent on edge weight than simply the presence/absence of an edge. The distribution of $s$ across all Event Reactions is shown in Figure \ref{fig:SSD}. We observe a range of behaviours; a prominent mode with relatively low structural similarity (i.e. page views are important), a broader intermediate mode (page views have some effect), and finally the sharp mode around $s=1$ (page views have little to no effect).

\subsection{Higher-level Topics of Attention}

Over all events, we now have a collection of Event Reactions. Many of these will be related through covering different stages of the same continuous event (e.g. different rounds of the FIFA World Cup), or through the re-emergence of events and news topics in time (e.g. updates about the Mueller investigation, or new natural disasters). We now turn to \textbf{RQb:} \textit{Are these groups of articles from different events related and do they form coherent topics?} We seek to identify the recurring groups of Wikipedia articles associated with news events---the \emph{topics} that are represented. Event Reactions from different events that are made up of broadly the same collection of Wikipedia articles are representative of a wider concept receiving repeated news exposure. We look to quantify the similarity between Event Reactions and use this to find the more closely related groups that represent Topics of Attention. We construct a higher-level network $H(V, E)$ of all Event Reactions (nodes). Edge weights are set as the weighted Jaccard similarity \cite{ioffe2010improved} between the sets of articles of each Event Reaction, weighted by their PageRank centrality in their respective networks \cite{page1999pagerank}, indicating similarity in content (and weighting more important articles to the concept more highly). This network contains all recorded instances of Event Reactions in the sample, representing their relation to one another over the course of one year. In order to identify the Topics of Attention (groups of related Event Reactions) We run a further stage of community detection over this network $H(V, E)$, using the Leiden algorithm with the Constant Potts Model as before. Whilst the nodes in the network represent snapshots of events centred on different points in time, $H(V,E)$ is not a temporal network. The resolution parameter is set at $r=0.067$, according to the robustness test set out in Appendix \ref{appendix:robustness2}. This process yields a partition of communities that are the Topics of Attention which we go on to label, validate, and explore.

\subsection{Topic Labelling and Validation}

In line with literature on news values and newsworthiness \cite{galtung1965structure, shoemaker1987deviance, harcup2001news, harcup2017news}, the Topics of Attention were sorted by several features detailed in Table \ref{tab:features}, with the top topics across each feature manually labelled. Several of these are based on the constructed time series, $W_{ij}$ for each Event Reaction ($R_{ij}$). This is a sum of the daily page views to each article $k$ in an Event Reaction ($p_k(t)$), weighted by their PageRank centrality ($w_k$) to the network they form;
\begin{equation}
    W_{ij}(t) = \sum_{k\in R_{ij}} w_k p_k(t).
    \label{eq:Wsumsc1}
\end{equation}
The time series is then centred in time to the max value occurring $\pm 1$ day from the recorded date of the event. For a Topic of Attention $A_\alpha$ with constituent Event Reactions $R_i$ ($i$ acting as an index for the Event Reactions in $A_\alpha$ and no longer referring to a specific event) and their associated time series $W_i$, the average prominence, magnitude, and deviance of a topic are then accordingly
\begin{equation}
    \text{Prominence}_\alpha = \frac{\sum_{W_i \in A_\alpha}{\text{median}(W_i(-30, -29, ..., 0))}}{\|A_\alpha\|},
\end{equation}
\begin{equation}
    \text{Magnitude}_\alpha =  \frac{\sum_{W_i \in A_\alpha}{W_i(0) - \text{median}(W_i(-30, -29, ..., 0))}}{\|A_\alpha\|},
\end{equation}
\begin{equation}
    \text{Deviance}_\alpha =  \frac{\sum_{W_i \in A_\alpha}\frac{W_i(0) - \text{median}(W_i(-30, -29, ..., 0))}{\text{median}(W_i(-30, -29, ..., 0))}}{\|A_\alpha\|}.
\end{equation}
Intuitively, prominence corresponds to how popular the subject(s) of an event or topic are before a particular event takes place. Magnitude corresponds to the absolute attention towards a given event when it does occur. Finally, deviance corresponds to unexpected event popularity, or how shocking a given event is, relative to its typically fairly unpopular subject matter. 

\begin{table}
\centering
\caption{A summary of the features with which we sort and examine the Topics of Attention.}
\begin{tabular}{p{4cm}|p{6cm}}
Feature                     & Description                                                                                                    \\   \hline
Number of associated events & Topics most frequently featured in the news (often background topics)                                          \\
Prominence                  & Topics which on average have the largest level of pre-existing attention, median page views (often background topics)     \\
Magnitude                   & Topics which on average receive the largest increase in page views when in the news                                \\
Deviance                    & Topics which on average receive the largest increase in page views, relative to their prominence, when in the news
\end{tabular}
\label{tab:features}
\end{table}

Two coders independently manually labelled a subset of 65 of the Topics of Attention by examining the constituent Wikipedia articles for each topic and the news events most associated with them (five events initially, with the option to see more). The labelling interface is shown in Appendix \ref{appendix:interface}. This set of topics were selected by taking the top 20 topics across each feature in Table \ref{tab:features} (some topics appear more than once across the four features, hence the total less than 80). Each coder was then presented with the combined list of labels and independently tasked with identifying where there was `strong agreement', `partial agreement', or `weak/no agreement' between labels. For the topics, 72.3\% of labels were in unanimous strong agreement, 7.7\% in strong-partial agreement (i.e., one coder ranked as strong agreement, and one as partial agreement), 15.4\% in unanimous partial agreement, and 4.6\% in weak/no agreement. The procedure demonstrates validity of the interpretable topics. For the purpose of display in figures and tables, in cases where there was not unanimous strong agreement between coders the first coder's labels are used.
\begin{table*}[]
\caption[Top topics by certain measures (minimum 10 events).]{Top topics by certain measures (min 10 events). Colour indicates quartile of structural similarity score, from red=bottom quartile to green=top quartile. Symbol indicates labelling agreement. No symbol: Unanimous Strong, *: Strong-Partial, **: Unanimous Partial, \dag: No agreement.}
\resizebox{\textwidth}{!}{
\begin{tabular}{l|llll}
 &
  \# Event Reactions &
  Prominence &
  Magnitude &
  Deviance \\ \hline
0 &
  \cellcolor{green}Countries &
  \cellcolor{orange}FIFA World Cup &
  \cellcolor{green}British Royal Family &
  \cellcolor{red}US administration \\
1 &
  \cellcolor{green}US Politics** &
  \cellcolor{red}Unknown\dag &
  \cellcolor{orange}FIFA World Cup &
  \cellcolor{yellow}California Wildfires 2 \\
2 &
  \cellcolor{yellow}US Domestic** &
  \cellcolor{green}British Royal Family &
  \cellcolor{orange}American Football &
  \cellcolor{yellow}Russia-Ukraine relations** \\
3 &
  \cellcolor{orange}Israel Palestine conflict &
  \cellcolor{green}Countries &
  \cellcolor{red}US administration &
  \cellcolor{yellow}Australian Prime Ministers** \\
4 &
  \cellcolor{orange}Syrian Civil War &
  \cellcolor{red}East Asia** &
  \cellcolor{yellow}Tennis &
  \cellcolor{orange}Wildfires \\
5 &
  \cellcolor{orange}Global Cities &
  \cellcolor{red}US Finance\dag &
  \cellcolor{yellow}Indian Politics &
  \cellcolor{orange}Turkey–United States relations** \\
6 &
  \cellcolor{orange}Middle East &
  \cellcolor{red}Unknown\dag &
  \cellcolor{green}Brett Kavanaugh Supreme Court nomination** &
  \cellcolor{orange}California Wildfires 1 \\
7 &
  \cellcolor{green}North Korea - South Korea relations* &
  \cellcolor{red}Trump Family &
  \cellcolor{red}Malaysian politics &
  \cellcolor{yellow}Australian Politics \\
8 &
  \cellcolor{orange}Terrorism in Afghanistan &
  \cellcolor{green}US Politics** &
  \cellcolor{yellow}Thai cave rescue &
  \cellcolor{green}White Helmets (Syrian civil war) \\
9 &
  \cellcolor{red}Trump Family &
  \cellcolor{yellow}Cryptocurrency &
  \cellcolor{yellow}California Wildfires 2 &
  \cellcolor{red}Florida \\
10 &
  \cellcolor{orange}Mueller Investigation &
  \cellcolor{green}Winter Olympics &
  \cellcolor{green}Gun violence in the United States &
  \cellcolor{yellow}Peruvian Politics \\
11 &
  \cellcolor{red}Putin \& Russian Politics &
  \cellcolor{orange}Big Tech &
  \cellcolor{yellow}Malaysian Politics &
  \cellcolor{green}Israel in Syrian civil war* \\
12 &
  \cellcolor{red}US political houses &
  \cellcolor{red}(South)eastern Europe** &
  \cellcolor{yellow}Musk \& Tesla &
  \cellcolor{orange}Puna Volcano \\
13 &
  \cellcolor{red}Korean conflict* &
  \cellcolor{red}US Presidency* &
  \cellcolor{red}US political houses &
  \cellcolor{red}Malaysian politics \\
14 &
  \cellcolor{green}Global Currencies &
  \cellcolor{yellow}Social Media &
  \cellcolor{red}Foundation of USA** &
  \cellcolor{green}Costa Rican politics \\
15 &
  \cellcolor{yellow}Latin America &
  \cellcolor{red}Foundation of USA** &
  \cellcolor{green}Tropical Storms &
  \cellcolor{yellow}Thai cave rescue \\
16 &
  \cellcolor{yellow}Yemeni Civil War &
  \cellcolor{red}Colombian drug trade* &
  \cellcolor{green}Toy retailers** &
  \cellcolor{green}Toy retailers** \\
17 &
  \cellcolor{green}Catalan Independence Movement &
  \cellcolor{yellow}US Domestic** &
  \cellcolor{yellow}Australian Politics &
  \cellcolor{green}Yemeni politics \\
18 &
  \cellcolor{orange}Saudi Royal Family &
  \cellcolor{orange}Global Cities &
  \cellcolor{yellow}US Gun violence &
  \cellcolor{yellow}Iranian Politics \\
19 &
  \cellcolor{orange}UK Politics &
  \cellcolor{orange}US military &
  \cellcolor{yellow}Olympics &
  \cellcolor{yellow}Georgian politics \\
20 &
  \cellcolor{green}Tropical Storms &
  \cellcolor{orange}Middle East &
  \cellcolor{green}NHL &
  \cellcolor{green}Johnson Family \\
\end{tabular}}
\label{tab:ToA}
\end{table*}

\section{Results and Discussion}

Studying the contents of the emergent Topics of Attention in Table \ref{tab:ToA} reveals various interesting details on current events as recorded on Wikipedia. Identified features include; a background concept space, strong geographical effects (including a heavy Anglosphere/US focussed bias), a focus on individuals, and breakout subtopics.

\subsubsection*{Background concept space}
Several of the top Topics of Attention by number of associated events (Countries, Global Cities, Tropical Storms, etc.) are those of lasting historical context. These topics also typically have high structural similarity---attention towards the topic is correlated with its structural composition on Wikipedia. Whilst the Event Networks are sampled from a current events records, much of the related content is built on and widely considered as part of long established knowledge. This supports the case of news events contributing to longer term narratives.

\subsubsection*{Strong geographical effects}
The Topics of Attention are strongly characterised by geography. Many of the labelled topics are specified by the region they are relevant to. It is also clear that when incorporating the structure of the knowledge graph and attention that many of the most prominent topics are Anglosphere focussed. This is of course partly a consequence of studying the English Wikipedia. However, the current events portal's nominal aim, and that of English Wikipedia as a whole, is to objectively cover global events and knowledge---something it still falls short on. Topics relating to the US and UK are covered with far higher granularity than those relating to other countries. That is, there are several top topics related to the intricacies of US politics, yet other countries typically have all related news summarised within a single topic. Considering the topic labels with some geographic link, 46\% are Anglosphere based (primarily US, with a handful UK and Australia based). This is not entirely unsurprising, given prior work on Wikipedia biases \cite{graham2014uneven, callahan2011cultural, hecht2009measuring, adams2019counts} as well as this work's focus on the English language Wikipedia. Nevertheless, this further validates assertions that rather than being the ``sum of all human knowledge'' \cite{slashdot_2004}, Wikipedia (in its various languages), through its content, structure, and access patterns is highly sensitive to its cultural setting. 

\subsubsection*{Focus on individuals}
Several of the labelled topics are focussed on, or strongly feature a powerful individual (e.g. Trump family, Putin \& Russian Politics, Musk and Tesla). This points to an audience sensitivity towards people that can be related to or reviled and is reflective of findings on the news values of celebrity/power elite \cite{harcup2017news}.

\subsubsection*{Breakout subtopics}
There are several cases where topics may be strongly related, yet one cluster achieves breakout popularity enough to distinguish itself from the original topic. These could correspond to the well studied phenomena of ``media storms'' \cite{boydstun2014two}, whereby there is intense media focus on a single issue. An example of this is the topic for North Korea - South Korea relations---representing an overview of related articles---and the Korean Conflict---which is the subject of more intense focus around events by the audience as indicated by the differing structural similarity scores. On top of the differing structural similarity scores, the Korean Conflict topic has higher prominence, magnitude, and deviance than the North Korea - South Korea relations topic. Another example is the broader US Politics topic compared to the US political houses or current US administration topic. The former represents stable knowledge attracting attention around the topic and the latter represent new, more unusual combinations of articles more closely associated with current events. This may be a consequence of the choice of Jaccard similarity for the higher-level graph edge weights, where news events create strong, synchronous deviations from typical page view behaviour across a very small group of articles that are still related to a wider group. Since the number of deviating articles for the individual event is small, the edge weight through Jaccard similarity to other related events with a larger set of articles is also small, leading to it not being included in the Topic of Attention. An alternative similarity metric such as the overlap (Szymkiewicz-Simpson) coefficient could account for this effect, though using this would likely smooth over any breakout clusters, interesting features unto themselves. 

\subsubsection*{Further remarks}

The qualitative discussion of results and exploration of content is important in contextualising findings in further works on Wikipedia's coverage of current events, as well as Wikipedia status as an `independent' data source for news media.  Beyond simply being indicators for the notable issues in the news over a year, the detected Topics of Attention and their properties are demonstrative of the assertion that there is a disconnect between the `editor's' Wikipedia and the `page viewer's' Wikipedia. The central tension of Wikipedia as both a slow moving encyclopaedic knowledge base and fast moving current events record is displayed in the ways the topics are constructed. In the first mode, the audience's access patterns align with the established structure of knowledge on Wikipedia. The truly interesting mode occurs when the audience's attention does not align with the article network and in effect establishes its own communities of related articles. This collective behaviour is what stretches Wikipedia both towards updating its content and remaining a popular information resource, and away from its traditional encyclopaedic grounding.

There are several limitations to the methods proposed here. Firstly is the issue of this being a single language study. There have been a number of articles on the varying content, coverage, and use of different language Wikipedias based on linguistic, cultural, and national focusses \cite{Aragon2011, Bao2012, hale2014multilinguals, lemmerich2019world}. One could contend that this means a single story from one community of people editing and viewing Wikipedia. A strong Anglosphere bias is indeed observed but we see it as the case the English Wikipedia is not the product of, nor the information tool, for a single, large, homogeneous community. \cite{welser2011finding, west2012data, yang2016did} all observe that certain editors occupy particular roles in lending substantive expertise towards particular categories, whether that be due to identity, education, or other personal interest, and the same would be expected of regular users (to some extent also supported by \cite{Singer2017}). In addition, whilst solely applied to English Wikipedia here, the majority of the methods used are language agnostic, and may be swiftly applied to other language Wikipedias, which may be a fruitful avenue to pursue.

The Current Events Portal is clearly not an exhaustive source of news stories, many of which would have no discernible effect on Wikipedia. Explicit editing guidelines state that ``Stories added to the main portal page should be of international interest'' \cite{wikipedia_2021CE}. Beyond this restriction, there are a very large number of people who regularly access information related to sports, entertainment, and popular culture, whose news stories are rarely featured on the current events portal. Celebrity deaths, for instance, have their own summary article, rather than residing on the current events portal. The topic map thus does not cover the universe of what might be considered news, and is sensitive to the contents of the news story source, the Wikipedia Current Events Portal, raising the issue of endogeneity. Sampling news events by `Wikifying' \cite{Milne2008} alternative sources such as news website RSS feeds would indeed yield a different set of events, though unfortunately due to editorial decisions, we of course arrive at a similar obstacle where there is no objective set of events.

A more thorough comparison between the topic landscape of several different news outlets would be of interest and an immediate application of the developed methods towards agenda-setting research, yet is outside the scope of this work. An argument in favour of choosing the Wikipedia Current Events Portal is that this collaborative recording of news is representative of the collective received importance of events, incorporating what the news recording and accessing communities consider relevant. This, together with time constraints and the simplicity of selecting descriptions already formatted with Wikipedia links, resulted in the decision being made to concentrate on the Current Events Portal.

\section{Conclusion}

The encyclopaedic origins of Wikipedia mean it is not set up as a ready-made data source for the study of news events. Long term information involving for each subject, involving many events, is aggregated in each Wikipedia article, as opposed to aggregating information about multiple subjects at the event level for each event (i.e. like a news website). Events, aside from in rare cases where a particularly notable event justifies its own article, and news topics do not have an established natural representation on Wikipedia. Equally, the broadly consistent, common structured information available to its huge audience, as well as how this audience accesses content on current events, is too appealing to ignore. In order to take advantage of this, one must establish a framework for event and topic level study. To this end, we have developed an approach for event sampling and topic detection on Wikipedia, with a focus on news topics, that takes into account article network structure, dynamics, and content.

The graph supported correlation network approach towards temporal community detection successfully detects stable Event Reactions, relating both short-term dynamics of attention through page views as well as long-term knowledge structures, thus addressing \textbf{RQa}. We have demonstrated its utility in identifying and exploring different Event Reactions, and in their aggregate how they represent Topics of Attention, the objective of \textbf{RQb}. These objects of study improve upon those used in prior work for their generality across topics, usage of the knowledge network rather than focus on individual articles, explicit relation to news events, incorporation of short and long term effects, and lack of reliance on detection through particular attention dynamics. The Topics of Attention on Wikipedia exhibit a background historical concept space, strong geographical effects, a focus on individuals, and breakout subtopics. The Topics of Attention (through their constuent Event Reactions) may also be resolved to volatile news topics and stable background topics. More importantly, they represent both facets of Wikipedia as a stable knowledge base and rapidly updating current events record.

Detecting Topics of Attention using Wikipedia has proven to be a non-trivial task. It is important to encapsulate the contrasting timescales of news and existing knowledge, the many to many relationship between events and topics, together with the corresponding dynamics of attention and memory building. Through this process we gain insight how topics are represented and accessed on Wikipedia and on which events are considered important enough to make it into the encyclopaedic record. Finally, generating Event Reactions, and wider Topics of Attention can enable more detailed event and topic level study in further work. Establishing representations of events and topics, beyond individual articles, on Wikipedia allows us to quantitatively address questions on theories of news media, collective memory, and historical recording in ways not previously possible without this kind of massive audience level data.

\backmatter

\bmhead{Data Availability Statement}

The code used and data generated during the current study are available in the WikiNewsTopics GitHub repository: \url{https://github.com/pgilders/WikiNewsNetwork-01-WikiNewsTopics}.




\section*{Statements and Declarations}

The authors declare no conflicts of interest concerning this piece of research.









\newpage

\begin{appendices}

\setcounter{figure}{4}                       
\renewcommand\thefigure{\arabic{figure}}   

\section{Wikipedia Data}
\label{appendix:data}

Here we provide some additional details on the data as described in Section \ref{sec:data}.

\subsection{Current Events Portal}

Entries in the Current Events Portal are sorted in 10 categories (Armed conflicts and attacks, Law and crime, Arts and culture, Politics and elections, Business and economy, Science and technology, International relations, Sports, Health and medicine, Disasters and accidents). The criteria for an event's appearance on the page is simply that the community believe it is a significant story, subject to guiding principles such as the length and depth of existing news coverage, though ultimately, decisions are based on consensus of editors on the individual merits of events \cite{wikipedia_2021CE, wikipedia_2021ITN}.

\subsection{Clickstream Networks}

Visits from popular sources outside Wikipedia are also recorded in the raw clickstream data, although we only include hyperlinks between Wikipedia articles, excluding links from external to Wikipedia and from Wikipedia's Main Page. In addition, in the raw data only links with $>10$ clicks over the course of each month are supplied. A comprehensive analysis of all of Wikipedia's historical links would require examining full content dumps or revision histories of every article. The authors of this study believe that clickstream data provides a more efficient means of identifying the most relevant links and outweighs any tradeoff in completeness. The clickstream data is used since firstly it offers a fast, reliable snapshot of the past article network structure of Wikipedia (compared to the time taken wrangling HTML from the complete Wikipedia dumps, or from individual article revisions through API calls), and secondly the weighting allows for a cutoff for spurious links.

\subsection{Page View Time Series}

The page view data is grouped into monthly datasets with hourly granularity for each Wikimedia project. We focussed on the page views towards articles in the English Wikipedia, which are reported under the prefixes for desktop Wikipedia (en.z), the mobile Wikipedia (en.m), and Wikipedia Zero (en.zero) (a discontinued project where Wikipedia access was available for free in developing countries which in practice accounts for relatively few views).

\subsection{Redirects}

A somewhat minor, yet often overlooked, facet of research on Wikipedia is the issue of page redirects \cite{wikipedia_2021RD}. A redirect is a page which automatically sends visitors to another page with the `correct' title, for example searching for `USA' or clicking a wikilink titled `USA' automatically redirects the user to the page `United States'. Redirects are very important since users and editors constantly search for articles, click on links, and edit links with alternative/abbreviated/previous names. In fact, it has been estimated up to 55\% of the articles in the main namespace of Wikipedia are `redirects' \cite{hill2014consider}. In order to ensure information for identical articles in the data is not being duplicated, redirects need to be resolved. Redirects are already resolved in the clickstream dataset (though pages may change in name over time e.g. `Meghan Markle' to `Meghan, Duchess of Sussex') but they are unresolved in the page view dataset, i.e. hits on redirect pages are recorded separately from hits for the true page name \cite{wikipedia_2021PVRD}.

\section{Community Detection Resolution Selection}
\label{appendix:robustness1}

\begin{figure}[h]
    \centering
    \includegraphics[width=0.75\textwidth]{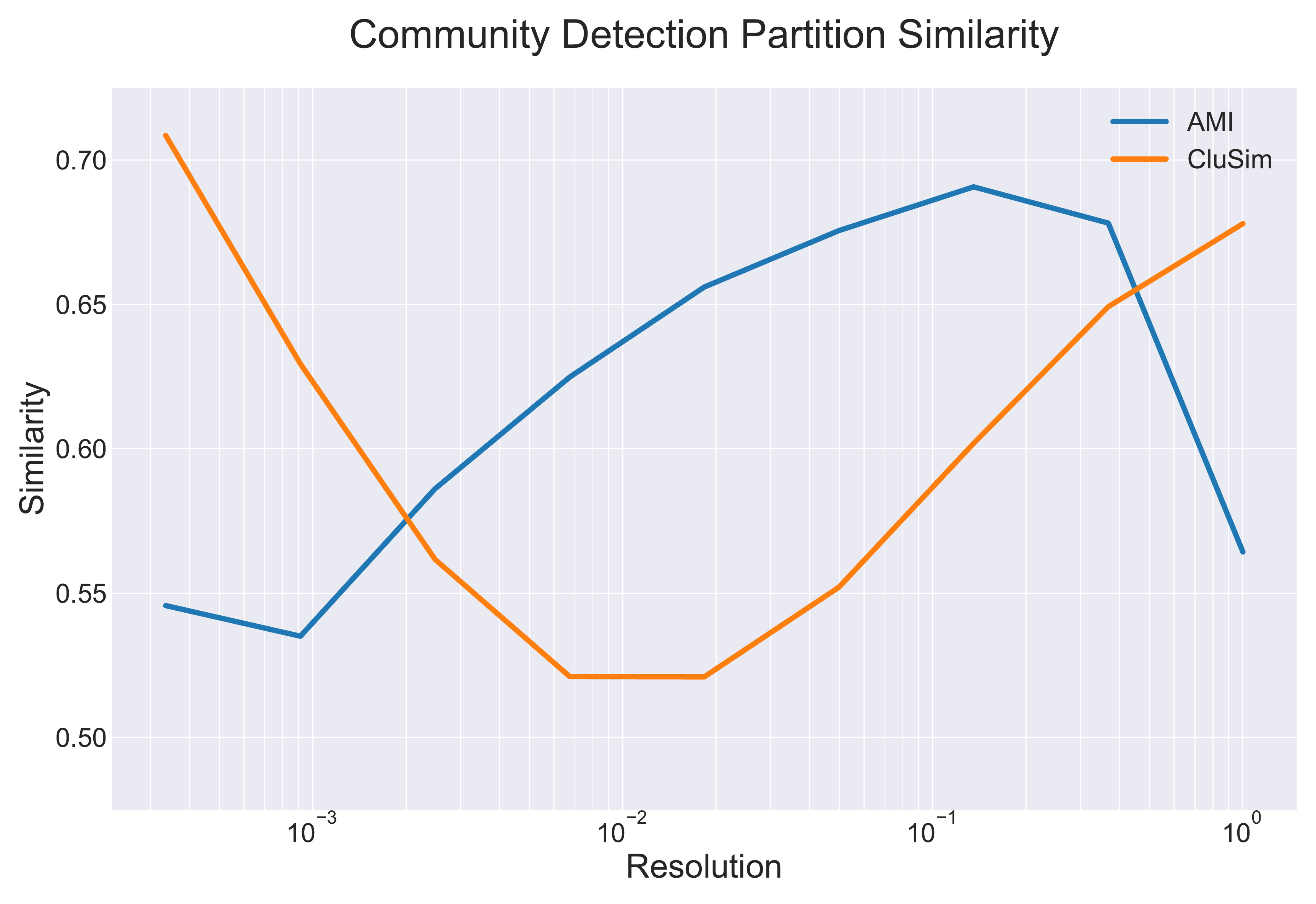}
    \caption[Partition stability for temporal community detection]{Partition stability. Partitions at smaller resolutions tend towards a single community and partitions at larger resolutions tend towards a unique community for each node. There are maxima for each similarity measure around $r=0.25$.}
    \label{fig:partstab}
\end{figure}

We perform a robustness test during the initial stage of temporal community detection on the Event Networks ($G_i$) to identify the resolution parameter value which gives stable, meaningful partitions. We test a random sample of 100 events, repeating the community detection process over a (logarithmic) range of resolution parameters, and compare the similarity in the obtained partitions. Similarity between partition $n$ and $n-1$ is calculated according to adjusted mutual information and CluSim element-centric similarity \cite{gates2019clusim} and shown in Figure \ref{fig:partstab}. Based on this test, the resolution parameter for further analysis is set according to the peak similarity around $r=0.25$.

\begin{figure}[h]
    \centering
    \includegraphics[width=0.75\textwidth]{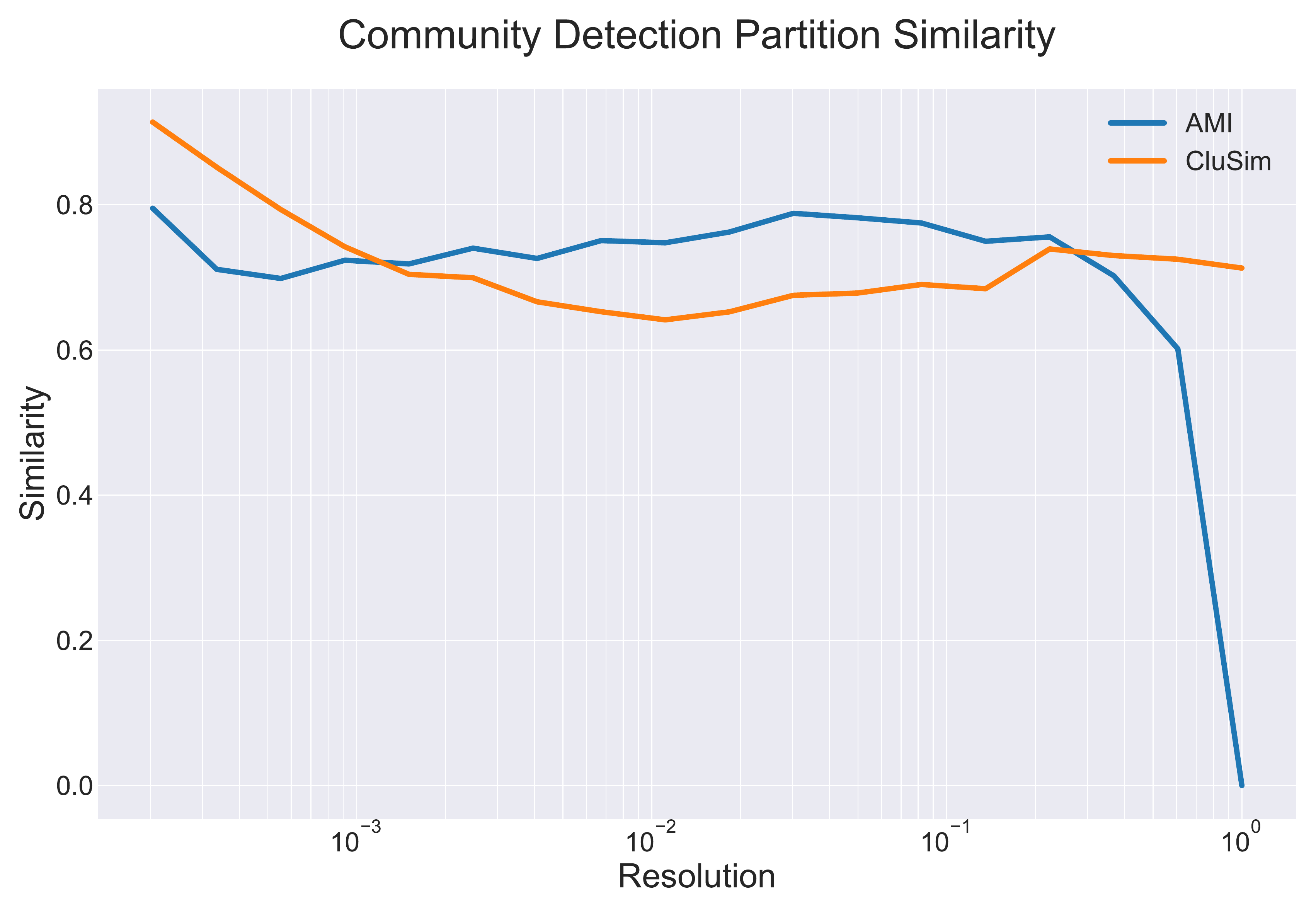}
    \caption[Partition stability for structural community detection]{Partition stability for static, unweighted community detection. Partitions at smaller resolutions tend towards a single community and partitions at larger resolutions tend towards a unique community for each node. Maximal similarity by AMI motivates selection of $r=0.030$ going forward.}
    \label{fig:partstab_str}
\end{figure}

\begin{figure}[h]
    \centering
    \includegraphics[width=0.75\textwidth]{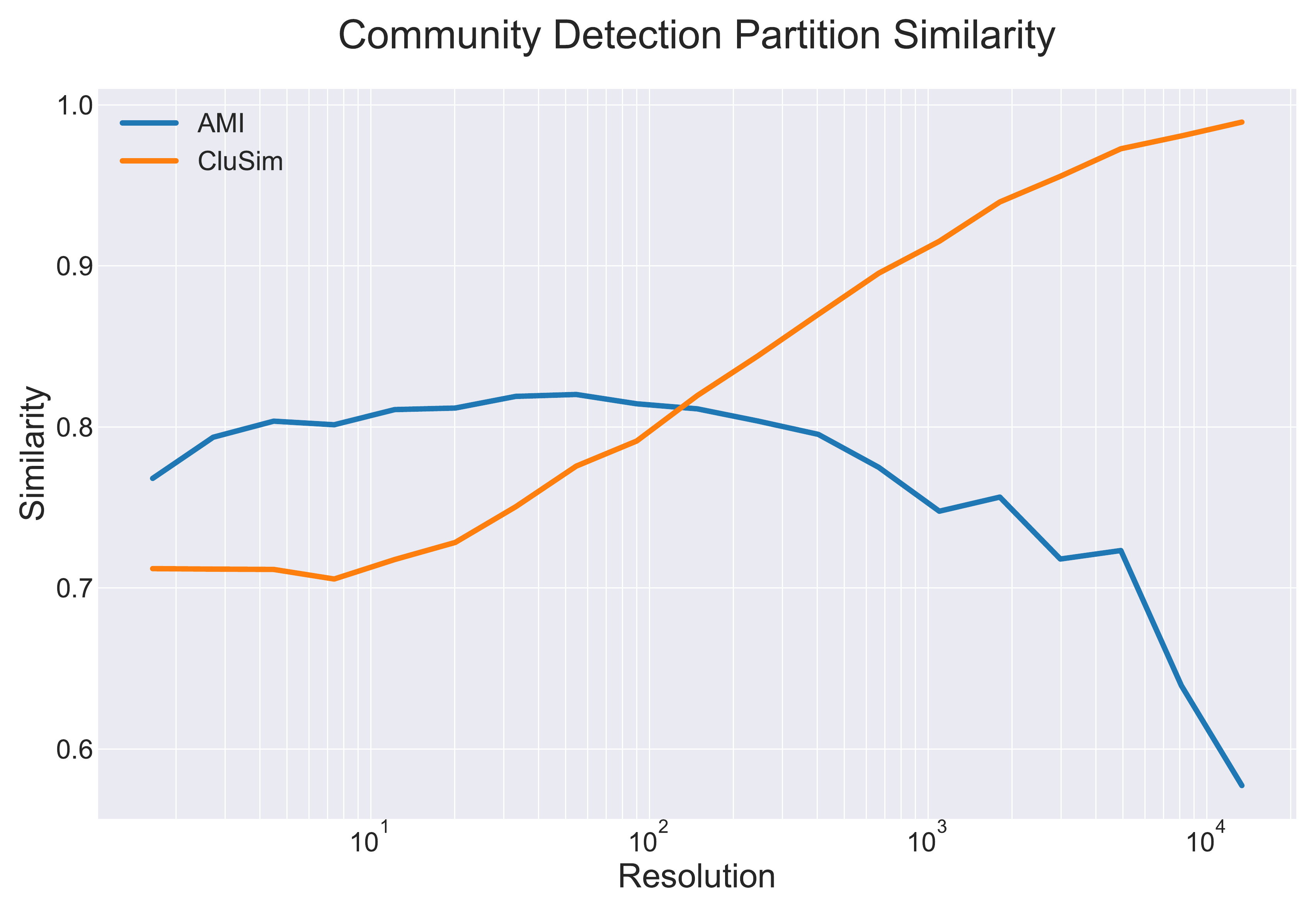}
    \caption[Partition stability for navigational community detection]{Partition stability for static, weighted community detection. Partitions at smaller resolutions tend towards a single community and partitions at larger resolutions tend towards a unique community for each node. Maximal similarity by AMI motivates selection of $r=54.6$ going forward.}
    \label{fig:partstab_nav}
\end{figure}

Similarly, in comparing to methods that only use structural or navigational info (Section \ref{sec:cdc}), we perform equivalent resolution tests for community detection approaches that use static networks based on the article hyperlink network and the aggregate clickstream navigation counts. Partition similarities for these approaches over a range of resolutions are shown in Figures \ref{fig:partstab_str} and \ref{fig:partstab_nav}. For further analysis with these approaches we then choose to use resolutions of 0.030 and 54.6 respectively.

\section{Higher-Level Network Community Detection}
\label{appendix:robustness2}

\begin{figure}[]
    \centering
    \includegraphics[width=0.75\textwidth]{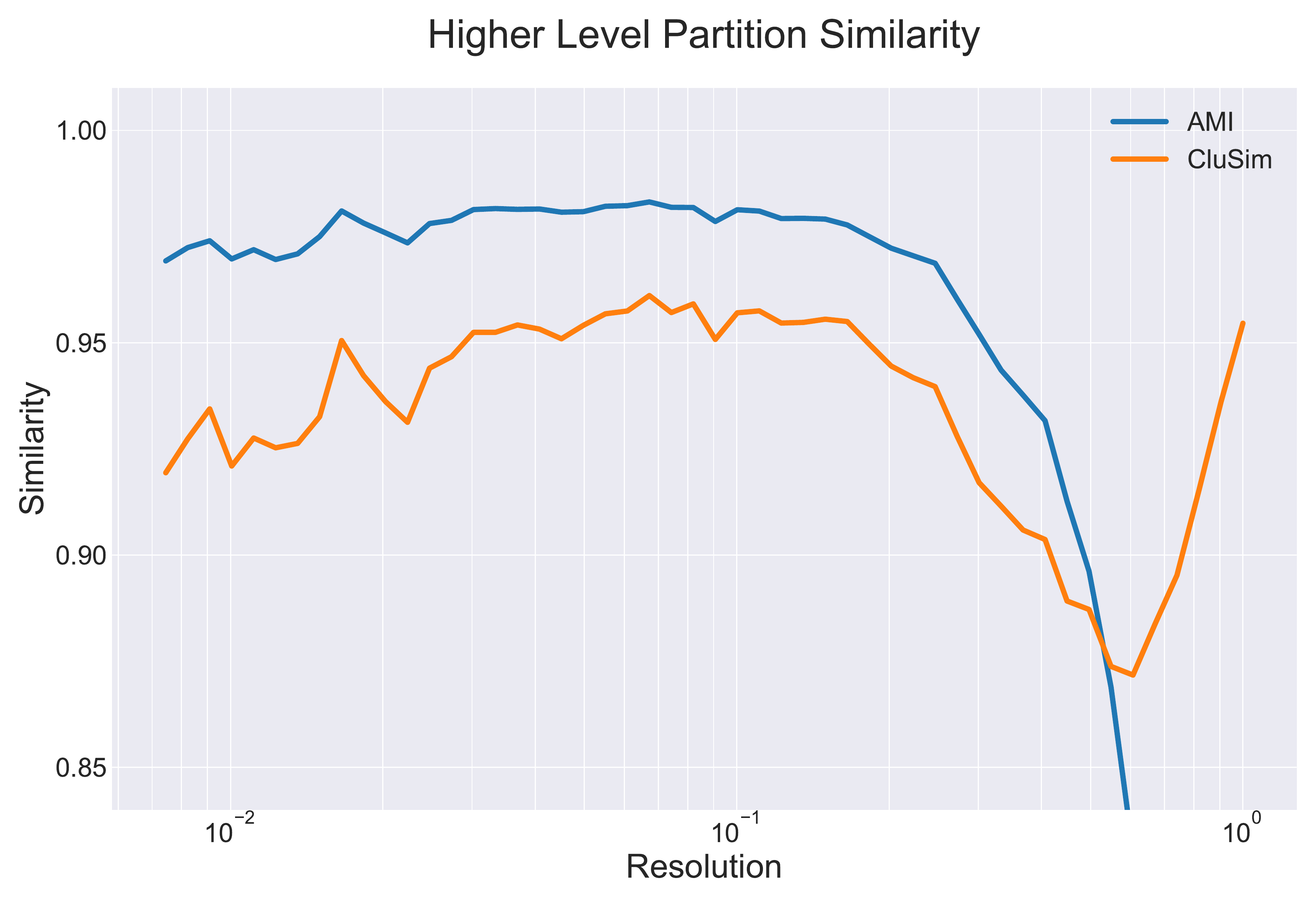}
    \caption[Higher-level network partition stability with varying resolution.]{Higher-level network partition stability with varying resolution. Partitions at smaller resolutions than those shown in the figure tend towards a single community and partitions at larger resolutions tend towards a unique community for each node. There are maxima for each similarity measure around $r=0.067$.}
    \label{fig:hpartstab}
\end{figure}

The approach of comparing partitions across a range of resolutions in \ref{appendix:robustness1} is repeated for community detection on the full higher-level network $H$ and shown in Figure \ref{fig:hpartstab}. In this case, the partition from the maxima around $r=0.067$ is chosen for further analysis.

\section{Topic Labelling}
\label{appendix:interface}

\begin{figure}[]
\centering
\includegraphics[width=\textwidth]{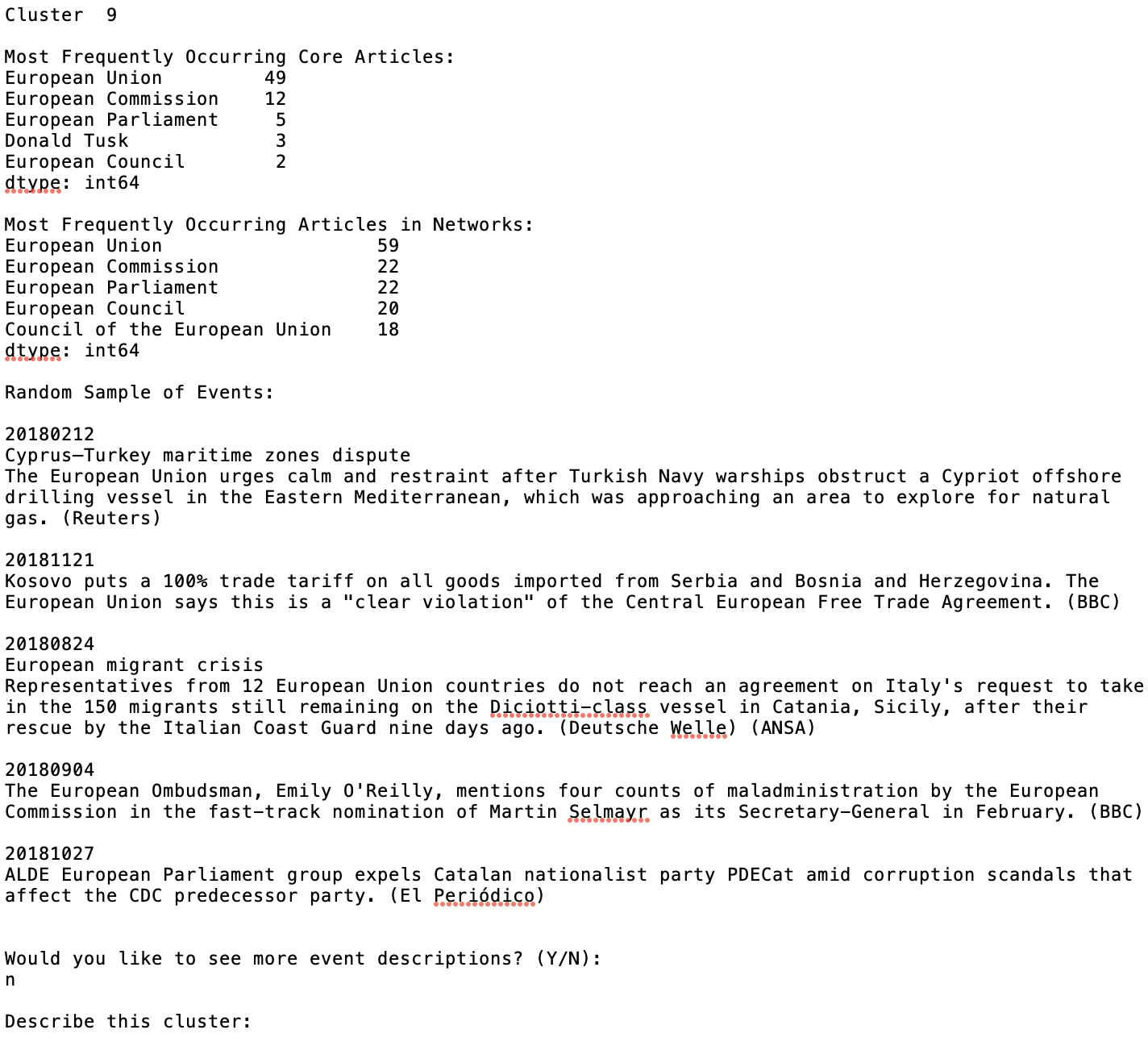}
\caption{The interface for labelling Topics of Attention, showing the most frequently occurring core articles, regular articles, and a sample of related events.}
\label{fig:interface}
\end{figure}

Figure \ref{fig:interface} shows the interface used by the independent labellers.

\end{appendices}

\clearpage 
\bibliography{sn-bibliography}


\end{document}